\documentstyle[12pt]{article}

\def\beq{\begin{equation}}
\def\eeq{\end{equation}}
\def\beqn{\begin{eqnarray}}
\def\eeqn{\end{eqnarray}}

\begin{document}
\begin{titlepage}

\begin{flushright}
ITEP-TH- 45/13\\
%December 23 2013/Draft
\end{flushright}

\vspace{1cm}

\begin{center}
{  \Large \bf Atomic collapse in graphene and cyclic RG flow}
\end{center}
\vspace{1mm}

\begin{center}

 {\large
 \bf   A.Gorsky$\,^{a,b}$ and F.Popov$\,^{a,b}$}

\vspace{3mm}

$^a$
{\it Institute of Theoretical and Experimental Physics,
Moscow 117218, Russia}\\[1mm]
$^b$
{\it Moscow Institute of Physics and Technology,
Dolgoprudny 141700, Russia}
\end{center}

\centerline{\small\tt gorsky@itep.ru }
\centerline{\small\tt popov@itep.ru }

\vspace{3cm}

\centerline
{\large\bf Abstract}
  \vspace{1cm}
In this Letter we consider the problem of screening of  external charge
in  graphene from the cyclic RG flow viewpoint.
The analogy with  conformal Calogero model is used
to suggest the interpretation of the  tower of
resonant states as tower of Efimov states.

%\end{center}

\end{titlepage}

\section{Introduction}
It is usually assumed that the RG flow connects
fixed points, starting at a UV repelling point and terminating at a IR attracting point. However it turned out that this open
RG trajectory does not exhaust all possibilities and
the clear-cut quantum
mechanical example of the nontrivial  RG limit cycle has been found in \cite{glazek}
confirming the earlier expectations. This finding triggered the search
for another examples of this phenomena which was quite successful.
The explicit examples
have been identified both in the systems with finite number  degrees of freedom \cite{ braaten,bavin,beane, sierra}
and in the field theory framework \cite{leclair, son,gorsky}.
It was also realized later  that the
Efimov states predicted long time ago for the three-body system  in the
context of nuclear physics with
the Efimov scaling of the energy levels are just the manifestation of this 
quite general phenomena.
The review on RG interpretation of the Efimov
phenomena can be found in \cite{hammer}.

The very phenomenon of the cyclic RG flow has been interpreted in \cite{son} as a kind
of generalization of the BKT-like phase transitions in two dimensions. One can start with a usual pattern of an RG flow connecting UV and IR fixed points and then consider a motion in a parameter space which results in a merging of the fixed points.  It was argued that when the parameter goes into the complex plane the cyclic behavior of the RG flow gets manifested and a gap in the spectrum arises.
This happens similar to the BKT phase
transition   when the deconfinement of vortices occurs at the critical temperature
and the conformal symmetry is restored at  lower temperatures.
The appearance of the RG cycles can be also interpreted in terms of the  peculiar
anomaly in the  classical conformal group \cite{anomaly}. This anomaly
has the origin in a kind of "falling to the center" UV phenomena which
could have quite different reincarnations.
Emphasize one more generic feature of this phenomena ---
the cyclic RG usually occurs in the system with at least two couplings. One of them
undergoes the RG cyclic flow while the second determines the period of the cycle.

The RG cycles have been
found in the non-relativistic Calogero-like model with $\frac{1}{r^2}$
potential   which enjoys the  conformal symmetry \cite{braaten,bavin,beane}.
In Calogero model with attraction  some small distance boundary conditions
are imposed on the wave function.
It is usually assumed that the wave function with $E=0$ at large $r$ does not depend
on the cutoff at small $r$. This condition yields the equation for the parameter
of a cutoff. It has  multiple solutions which can be
interpreted in terms of  the tower of shallow bound states  with
the Efimov scaling in the regularized Calogero model with attraction.
The scaling factor is determined by the Calogero coupling constant.

In this Letter we shall consider the similar problem in 2+1 dimensions
which physically corresponds to the external charge in the graphene plane.
The problem has a classical conformal symmetry and is the relativistic  analogue
of the conformal non-relativistic Calogero-like system. Due to  conformal symmetry
we could expect the RG cycles and Efimov-like states in this problem
upon imposing the short distance cutoff.
The issue of the charge in the graphene plane has been discussed
theoretically \cite{levitov2,gra1,gra2} and experimentally \cite{exp1,exp2}. It was
argued that indeed there is the tower of "quasi-Rydberg" states with the exponential
scaling \cite{levitov}. The situation can be interpreted as an atomic collapse phenomena similar
to the instability of $Z>137$ superheavy atoms in QED \cite{atoms}. The possible role of the
conformal symmetry in the atomic collapse problem in graphene
has been mentioned in \cite{levitovtalk}.

We shall perform the regularization analysis in  2+1 case
similar to Calogero model and identify the Efimov-like states
from this perspective. The period of the RG cycle will be
identified with the charge value  and the 2+1 relativistic
analog of the anomaly in the commutator has been considered.

The Letter is organized as follows. First, we will briefly review the
RG picture for non-relativistic Calogero model . In Section 3 we will perform
the  RG analysis imposing the small distance cutoff. In Section 4 we shall consider
the relativistic version of the anomaly in the commutator. Some
comments on the results obtained can be found in the Conclusion.

\section{RG cycles in nonrelativistic quantum mechanics}

In this Section we consider the example of the RG limit cycle
in the non-relativistic system with  the attractive $\frac{1}{r^2}$ Calogero potential regularized at short distances
in some way. Two most popular regularizations involve
the square-well
potential \cite{bavin,beane} or the delta-shell potential \cite{braaten}.
In both cases the depth of the potential is logarithmic function of the short
distance cutoff.

Consider the particle in the attractive potential
\beq
V(r)= -(1/4 +\mu^2)r^{-2} , r>R
\eeq
\beq
V_{short} = const , r<R
\eeq
It turns out that the
potential regularized at short distances has infinitely many shallow S-wave bound states.
They accumulates at
the threshold energy $E=0$. Near the threshold the ratio of the energies of the
successive states approaches $e^{\frac{2\pi}{\mu}}$ and the bound state
energies behave as
\beq
E_n\rightarrow ce^{-\frac{2\pi}{\mu}(n-n_0)}
\eeq
The model before regularization respects the conformal
symmetry which is  broken by the boundary conditions
imposed by the short-distance cutoff.

The RG flow is formulated in terms of the tuning parameter $\lambda$
which enters the short distance potential
\beq
V(r,\lambda)= - \frac{\lambda}{R^2} \qquad r<R
\eeq
for the "spherical shell" potential. We assume that $\lambda$ is R dependent
function. The dependence on the RG scale is fixed by some  condition introduced
by hands. The most convenient RG condition is the requirement that the regularized
potential reproduces the $E=0$ wave function at $r>R$. It can be shown that
the shallow bound states are reproduced as well in the leading approximation.
Note that we could require the RG condition not only for zero energy state
but for any shallow state. The delta-shell regularization is defined by the
following cutoff potential
\beq
V_{delta}= - \frac{\lambda}{R}\delta(r-R_{-}) \qquad r\leq R
\eeq
where $R_{-}$ is close to R but lies inside the interval $0<r<R$.

In the spherical square wall the RG condition implies the  equation
\beq
\sqrt{\lambda}\cot(\sqrt{\lambda}) = \frac{1}{2} + \mu \cot(\mu \ln(R/R_0))
\eeq
which has infinitely many solutions. In the delta-well regularization
the  RG equation reads as
\beq
\lambda(R)= \frac{1}{2} -   \mu \cot(\mu \ln(R/R_0))
\eeq
In both cases the period of the RG cycle is defined by the Calogero
coupling constant and equals $\frac{2\pi}{\mu}$.

Therefore we see that in the conformal non-relativistic quantum mechanics
one can define the cyclic RG flow for the parameter of the cutoff potential.
At each cycle one bound state appears or disappears from the spectrum.

\section{RG cycle  in graphene}
In this Section we discuss the similar problem in 2+1 dimensions
which physically corresponds to the external charge in the graphene plane.
Consider  an electron in graphene which interacts with
an external charge. The two-dimensional Hamiltonian reads as,
\beq
H_D = v_{F} \sigma_i p^i - V(r), \qquad i=1,2.
\label{dirac_ham}
\eeq
where the external charge creates a Coulomb potential
\beq
V(r) = -\frac{\alpha}{r}, \qquad r\ge R.
\label{pot_coul}
\eeq

As we shall see, the solution in presence of the potential (\ref{pot_coul}) oscillates  at the origin and needs to be regularized by some cutoff $R$. Hence close enough to the origin $r\le R$ the potential (\ref{pot_coul}) gets replaced by some constant potential $V_{reg}(r,\lambda(R))$.
%When energy is enough small($|E| \ll \frac{\hbar^2}{2 m R^2}$).
The renormalization condition for the $\lambda$ parameter is that the zero-energy wave function is  independent on the short-distance regularization. This condition is chosen similarly to that of the renormalization in the Calogero system. Hence our primary task is to find the zero-energy solution to the Dirac equation,

\begin{equation}
    H_D\psi_0=0.
    \label{dirac}
\end{equation}

Since the Hamiltonian commutes with the $J_3$ operator,

\begin{equation}
    J_3=i\frac{\partial}{\partial \varphi}+\sigma_3, \qquad \left[ H_D, J_3 \right]=0,
    \label{J3}
\end{equation}
we can look for the solutions of (\ref{dirac}) in the form:

\begin{equation}
    \psi_0=\left( \begin{array}{l} \chi_0(r)\\ \xi_0(r)e^{i\varphi}  \end{array}\right), \qquad J_3\psi_0=\psi_0.
    \label{spinor}
\end{equation}

In polar coordinates the equation (\ref{dirac}) reads as:

\begin{equation}
    \left\{
        \begin{array}{l}
            -i\hbar v_F \left( \partial_r+\frac 1r \right)\xi_0  = -V(r) \chi_0,\\
            -i\hbar v_F \partial_r \chi_0= -V(r)\xi_0,
        \end{array}
    \right.
    \label{dirac1}
\end{equation}
which is equivalent to

\begin{equation}
    \left\{
    \begin{array}{l}
        \xi_0(r)=i\hbar v_F (V(r))^{-1}{\partial_r \chi_0},\\
        \partial_r^2 \chi_0+ \left( \frac{1}{r}-\frac{V'(r)}{V(r)} \right)\partial_r\chi_0 +\frac{V^2(r)}{\hbar^2 v_F^2} \chi_0 = 0.
    \end{array}
    \right.
    \label{Dechi}
\end{equation}

For the potential $V =- \frac{\alpha}{r}$ we get the following equation on $\chi_0(r)$:

\begin{equation}
    \partial_r^2\chi_0+\frac{2}{r}\partial_r\chi_0+\frac{\beta^2}{r^2}\chi_0=0, \qquad \beta=\frac{\alpha}{\hbar v_F}.
    \label{eq_coul}
\end{equation}
Supposing that $\beta^2 = \frac{1}{4} + \nu^2$ we write the solution as

\begin{equation}
    \chi_0= \sqrt{r} \left(c_- \left( \frac{r}{r_0} \right)^{-i\nu}+c_+ \left( \frac{r}{r_0} \right)^{i\nu}  \right)\propto \sqrt{r}\sin\left( \nu \log \frac{r}{r_0}+\varphi \right).
    \label{sol_coul}
\end{equation}
It shares the properties of the ground-state Calogero wave-function and at nonzero $c_{\pm}$  generates its own intrinsic length scale. In order to fix the  constant we need to introduce a cut-off potential:

\begin{equation}
    V(r)=\left\{
    \begin{array}{l}
        -\frac{\alpha}{r}, r>R,\\
        V_{reg}=-\hbar v_F \frac{\lambda}{R}, r\le R.
    \end{array}
    \right.
    \label{reg_pot}
\end{equation}

The dilatation  acts on $\chi$ as following:

\begin{equation}
    r\partial_r \chi_0=\left( \frac 12 +\nu \cot \left( \nu\log \frac{r}{r_0} \right) \right)\chi_0.
    \label{dil_coul}
\end{equation}

%For further purpose let us write down logarithmic derivative
%\[
%r \frac{\chi'}{\chi} = \frac{1}{2} + \nu \cot[\nu \log(\frac{\rho}{\rho_0})]
%\]
%\subsection{Constant potential}

For the constant potential $V_{reg}$ we get from $(\ref{Dechi})$:

\begin{equation}
    \partial^2_r \chi_0^{reg}+\frac{1}{r}\partial_r \chi_0^{reg} + \frac{\lambda^2}{R^2} \chi_0^{reg} = 0.
    \label{eq_const}
\end{equation}
and  solution of (\ref{eq_const})  regular at the origin is

\begin{equation}
    \chi_0^{reg}\propto J_0\left(\lambda \frac{r}{R} \right).
    \label{sol_const}
\end{equation}

Computing the action of the dilatation  on the solution at short distance and equating it to the action of the dilatation  (\ref{dil_coul}) we get the equation on the  parameter of regularization:

%\[
%\chi \propto J_0(\frac{|V|}{\hbar v_F} \rho),\rho \frac{\chi'}{\chi} = -\frac{|V| \rho}{\hbar v_F} \frac{J_1(\frac{|V|}{\hbar v_F} \rho)}{J_0(\frac{|V|}{\hbar v_F} \rho)}
%\]
%\subsection{Square-Well Regularization}
%Now we are ready to introduce  square-well regularization. Suppose that we choose $V_{reg} = - \lambda \frac{\hbar v_F}{R} ,\rho <R$. Now we are going to tune $\lambda$
%\[
%\frac{1}{2} +  \nu \cot[\nu \log(\frac{R}{\rho_0})] = - \lambda \frac{J_1(\lambda)}{J_0(\lambda)}
%\]

\begin{equation}
    \frac 12+ \nu\cot\left( \nu\log\left( \frac{R}{r_0} \right) \right)=-\lambda\frac{J_1(\lambda)}{J_0(\lambda)}.
    \label{rg_graph}
\end{equation}
The equation (\ref{rg_graph}) defines $\lambda$ as a multi-valued function of $R$. The period of the RG flow corresponds to jump from one branch of the $\lambda(R)$ function to another.

Now we shall derive the bound states in the (\ref{pot_coul}) potential and  consider  the Dirac equation,

\begin{equation}
    H_D\psi_\kappa=-\hbar v_F \kappa \psi_\kappa.
    \label{dirac_E}
\end{equation}
The equation on $\chi$ analogous to (\ref{Dechi}) reads as:

\begin{equation}
    \partial^2_r\chi_\kappa+\frac{2\beta-\kappa r}{\beta-\kappa r}\frac{1}{r}\partial_r\chi_\kappa+\left( \frac{\beta}{r}-\kappa \right)^2\chi_\kappa=0.
    \label{eq_coul_E}
\end{equation}
Asymptotically at $r\gg \frac{\beta}{\kappa}$ the solution to (\ref{eq_coul_E}) regular at infinity is given by the Hankel function,

\begin{equation}
    \chi_\kappa\propto H_0^{(1)}(i\kappa r).
    \label{sol_inf}
\end{equation}

At small $r\ll \frac{\beta}{\kappa}$ the solution is not regular at the origin,

\begin{equation}
    \chi_\kappa\propto \sqrt{r}\sin\left( \nu \log \frac{r}{r_0} \right),
    \label{sol_E}
\end{equation}
and we  need for the regulator potential. Solving  the Dirac equation (\ref{dirac_E}) in presence of the constant potential $V_{reg}$ and computing the action of the dilation operator,

\begin{equation}
    r\partial_r \chi_k^{reg}=-\left( \lambda-\kappa R \right)\frac{J_1\left(\lambda-\kappa R \right)}{J_0\left( \lambda-\kappa R \right)} \chi_\kappa^{reg},
    \label{dil_reg_E}
\end{equation}
we can equate (\ref{dil_reg_E}) to the action of the dilatation operator on (\ref{sol_E}) and get the equation on the spectrum of the bound states,

\begin{equation}
    \frac 12+\nu\cot\left( \nu\log\left( \kappa R \right) \right)=-\left( \lambda-\kappa R \right)\frac{J_1\left( \lambda-\kappa R \right)}{J_0\left( \lambda-\kappa R \right)}.
    \label{rg_kappa}
\end{equation}
This condition gives the spectrum of infinitely many shallow bound states,with Efimov scaling

\begin{equation}
    \kappa_n=\kappa_* \exp\left( -\frac{\pi n}{\nu} \right), \qquad \kappa\to \infty.
    \label{kappa}
\end{equation}

\section{Anomalous commutator algebra}
\label{sec:anomaly}
Let us make some comments on the algebraic counterpart
of the phenomena considered following \cite{anomaly}.
As we have mentioned above the
conformal symmetry is the main player since Hamiltonians under
consideration are  scale invariant before regularization.
Actually this group can be thought of as the example of
spectrum generating algebra when the Hamiltonian is identified with one of
the generators or is expressed in terms of the generators
in a simple manner. This situation is familiar from the exactly or
quasi-exactly solvable problems in quantum mechanics when the dimension
of the group representation selects the size of the algebraic part of the
spectrum.

Let us introduce the generators of the $SO(2,1)$ conformal
algebra $J_1, J_2, J_3$ :
the Calogero Hamiltonian,
\beq
J_1= H= p^2 +V(r),
\eeq
the dilatation 
\beq
J_2=D = t H -\frac{1}{2}(pr + r p),
\eeq
and the generator of special conformal transformations,
\beq
J_3=K =  t^2 H - \frac{t}{2}(pr +r p) +\frac{1}{2}r^2.
\eeq
They satisfy the relations of the ${\bf so}(2,1)$ algebra:
\beq
[J_2,J_1]= -iJ_1,\qquad [J_3,J_1]= -2iJ_2,\qquad [J_2,J_3]= iJ_3
\eeq

The singular behavior of the potential at the origin
amounts to a kind of anomaly in  the ${\bf so}(2,1)$ algebra \cite{anomaly},
\beq
A(r)=-i[D,H]+H,
\eeq
which in $d$ space dimensions can be presented in the
following form:
\beq
A(r)= -\frac{d-2}{2}V(r) + (\nabla_i r^i) V(r).
\label{simple}
\eeq
The simple arguments imply the following relation
\beq
\frac{d}{dt}\langle D\rangle_{\mathrm{ground}}=E_{\mathrm{ground}},
\label{anomaly}
\eeq
where the matrix element is taken over the
ground state.

It turns out that (\ref{anomaly}) is fulfilled
for the singular potentials in Calogero-like model
or in models with contact potential, $V(r)=g\delta(r)$. The expression for anomaly
does not depend on the regularization chosen.
Moreover the detailed analysis demonstrates that
the anomaly is proportional to the $\beta$-function
of the coupling providing the UV regularization
as can be expected.

The similar calculation of the anomaly for the 2+1
case can be performed for arbitrary state.
\beq
\left\langle{\frac{d D}{d t}}\right\rangle_\psi = {\Xi}_\psi =  - \int d^2 x \psi^*(V(x)+x_i \partial_i V(x)) \psi,
\eeq
which yields with square-well regularization
\beq
{\Xi}_\psi =\hbar v_F \frac{\lambda(R)}{R} \frac{\int \limits^{R}_0 r |\psi|^2 dr}{\int \limits^{\infty}_0 r |\psi|^2 dr}.
\eeq

Wave function for the very shallow state has the asymptotic behavior
\beq
\chi = \left\{
\begin{array}{l}
A r^\frac{1}{2}\sin(\nu \log(\kappa r)),L \gg r > R \\
B J_0(\lambda \frac{r}{R}),r<R
\end{array}
\right.
\eeq
\beq \phi  = \left\{
\begin{array}{l}
A \frac{\nu}{\beta} r^\frac{1}{2} \cos(\nu \log(\kappa r))+ A \frac{1}{2\beta} r^\frac{1}{2} \sin(\nu \log(\kappa r)) ,L \gg r > R \\
B J_1(\lambda \frac{r}{R}),r<R
\end{array}
\right.
\eeq
 Since function behaves well near the origin ($\sqrt{r} \sin(r) \to 0, r \to 0$ and $J_\alpha(r) \sim r^\alpha$) we get
\beq
\int \limits^\infty_0 |\psi|^2 r dr = A^2 N + O(R),R\to 0
\eeq

 Due to continuity we can use the following condition $\chi(R-0)=\chi(R+0),\phi(R-0) = \phi(R+0)$ and obtain
\beq
\Xi(R) = \lambda(R) \frac{\hbar v_F}{R A^2 N} \int \limits^R_0 |\psi|^2 r d r =  \frac{\hbar v_F}{N}(\frac{1}{2} + \frac{\nu^2}{2\beta^2}+\frac{1}{8\beta^2})\lambda(R)R^2
\eeq
Hence, if we choose $\log$-periodic behavior of $\lambda(R)$, we immediately get in the $R\to 0$,\quad $\Xi(R) \to 0$. On the other hand if we choose continuous branch of $\lambda(R)$ there is no a well-defined limit.

\section{Conclusion}

In this note we have interpreted the atomic collapse problem
considered in \cite{levitov} in terms of the RG cycle. The consideration
was  similar to the non-relativistic Calogero case.
Introducing the short distance cutoff and imposing the RG condition
in terms of  zero energy wave function  we get the RG equation
with limit cycle.
It was found that the period of the RG cycle  is fixed by the
value of the external charge. The spectrum of the states
with the Efimov scaling gets rearranged upon the cycle and
$E_{n+1}$ starts to  play the role of $E_n$.

It would be interesting to push forward the analogy with
$Z>137$ phenomena further. The possible conjecture could sound
as follows. When one discusses the Efimov state the general picture
involves the pair of particles near the threshold while
the third particle provides the
three-body bound state. We could imagine that in the atomic
collapse problem the electron pair plays the role
of two particles near threshold while the external source plays
the role of the third body.
It would be also interesting to discuss the possibility of the
similar cyclic RG flow in the bilyer graphene with
possible Mexican hat potential discussed in \cite{skinner}.

The work of A.G.  was supported in part
by grants RFBR-12-02-00284 and PICS-12-02-91052. A.G. thanks
FTPI at University of Minnesota where the part of this work has been done
for the hospitality and support.
We would like to thank K. Bulycheva, D. Kharzeev, L. Levitov, B. Shklovsky and
M. Voloshin for the useful discussions and comments.

\end{document}